\documentclass{nature_mod}

\usepackage{url}
\usepackage{graphicx}
\usepackage{aas_macros}
\usepackage{amssymb,amsmath}
\usepackage{multirow}

\renewcommand{\thefootnote}{\alph{footnote}}

\long\def\symbolfootnote[#1]#2{\begingroup%
\def\thefootnote{\fnsymbol{footnote}}\footnote[#1]{#2}\endgroup} 

\title{A close-pair binary in a distant triple supermassive black-hole system}

\author{R.~P.~Deane$^{1,2}$, Z.~Paragi$^3$, M.~J.~Jarvis$^{4,5}$, M.~Coriat$^{1,2}$, G.~Bernardi$^{2,6,7}$, R.~P.~Fender$^{4}$, S.~Frey$^8$, I.~Heywood$^{9,6}$, H.-R.~Kl\"ockner$^{10}$, K.~Grainge$^{11}$, C.~Rumsey$^{12}$}

\begin{document}

\maketitle

\noindent Published online by {\sl Nature} on 25 June 2014. DOI: 10.1038/nature13454

\begin{affiliations}

 \item Astrophysics, Cosmology and Gravity Centre, Department of Astronomy, University of Cape Town, Cape Town, South Africa
 \item Square Kilometre Array South Africa, Pinelands, Cape Town, South Africa
 \item Joint Institute for VLBI in Europe, Dwingeloo, The Netherlands
 \item Astrophysics, Department of Physics, University of Oxford, Oxford, UK
 \item Physics Department, University of the Western Cape, Belville, South Africa
 \item Centre for Radio Astronomy Techniques and Technologies, Department of Physics and Electronics, Rhodes University, Grahamstown, South Africa
 \item Harvard-Smithsonian Center for Astrophysics, Cambridge, MA, USA
 \item Satellite Geodetic Observatory, Institute of Geodesy, Cartography and Remote Sensing, Budapest, Hungary
 \item CSIRO Astronomy and Space Science, Marsfield, NSW, Australia
 \item Max-Planck-Institut f\"ur Radioastronomie, Bonn, Germany
 \item Jodrell Bank Centre for Astrophysics, School of Physics and Astronomy, The University of Manchester, Manchester, UK
 \item Astrophysics Group, Cavendish Laboratory, University of Cambridge, Cambridge, UK

\end{affiliations}

\begin{abstract}

Galaxies are believed to evolve through merging\cite{Springel2005}, which should lead to multiple supermassive black holes in some\cite{Begelman1980,Volonteri2003, Kulkarni2012}. There are four known triple black hole systems\cite{Tonry1984, Barth2008, Liu2011, Schawinski2011}, with the closest pair being 2.4~kiloparsecs apart (the third component is more distant at 3~kiloparsecs)\cite{Liu2011}, which is far from the gravitational sphere of influence of a black hole with mass $\sim$10$^9$~M$_\odot$ (about 100~parsecs). Previous searches for compact black hole systems concluded that they were rare\cite{Burke-Spolaor2011}, with the tightest binary system having a separation of 7 parsecs\cite{Rodriguez2006}. Here we report observations of a triple black hole system at redshift z=0.39, with the closest pair separated by $\sim$140 parsecs. The presence of the tight pair is imprinted onto the properties of the large-scale radio jets, as a rotationally-symmetric helical modulation, which provides a useful way to search for other tight pairs without needing extremely high resolution observations. As we found this tight pair after searching only six galaxies, we conclude that tight pairs are more common than hitherto believed, which is an important observational constraint for low-frequency gravitational wave experiments\cite{Wyithe2003, Sesana2013}.

\end{abstract}

SDSS\,J150243.09+111557.3 (J1502+1115 hereafter) was identified as a quasar at redshift $z = 0.39$ with double-peaked [OIII] emission lines, which can be a signature of dual active galactic nuclei (AGN)\cite{Smith2010}. Adaptive optics assisted near-infrared ({\sl K}-band) images reveal two components separated by 1.4 arcseconds (7.4 kpc) and are referred to as J1502P and J1502S\cite{Fu2011a}. Integral-field spectroscopy determined that the double-peaked [OIII] emission in the SDSS spectrum is explained by a 657~km\,s$^{-1}$ velocity offset between J1502S and J1502P, which are dust-obscured and unobscured AGN respectively. These two components were also observed as steep-spectrum radio sources ($\alpha < -0.8$, $S_\nu \propto \nu^{\alpha}$) with the Jansky Very Large Array (JVLA) at 1.4, 5, and 8 GHz. The combination of the above results supported the discovery of a kiloparsec-scale dual AGN system\cite{Fu2011b}.

We performed 1.7~GHz and 5~GHz Very Long Baseline Interferometry (VLBI) observations of J1502+1115 with the European VLBI Network (EVN), revealing two flat-spectrum ($\alpha \sim -0.1 \pm 0.1$) radio components within the spatially-unresolved near-infrared component J1502S. Flat radio spectra identify optically-thick radio emission which is characteristic of the core of relativistic jets generated by an accreting black hole\cite{Blandford1979}. The two components (labelled J1502SE and J1502SW in Fig.~1a) are each marginally resolved and have an angular separation of $\sim$26 milli-arcseconds which corresponds to a projected spatial separation of $\sim$138 parsec at $z=0.39$. Both J1502SE and J1502SW have radio luminosities of $L_{1.7} = 7 \times 10^{23}$~W\,Hz$^{-1}$ and brightness temperatures of $T_{\rm B} \ \gtrsim \ 2 \times 10^{8}$~K, consistent with actively accreting, intermediate radio luminosity nuclei. The high brightness temperatures, flat spectra, and co-spatial centroids (of both J1502SE and J1502SW) at both frequencies strongly support that these two radio components are associated with two separate, accreting supermassive black holes (SMBHs). Alternative scenarios are discussed and ruled out in the Methods section. Arcminute Microkelvin Imager (AMI) Large Array observations at 15.7 GHz suggest that the two radio cores (J1502SE/W) flatten the overall J1502S spectrum at higher frequency (see Fig.~4). The third active nucleus (J1502P) at a projected distance of 7.4 kpc from J1502S has a steep radio spectrum and is not detected in the VLBI observations. 

A re-analysis of archival JVLA observations provides supporting evidence for the inner binary discovery in J1502S. The JVLA 5~GHz map of J1502S and J1502P is shown in Fig.~1b, and Fig.~1c shows the point-source-subtracted JVLA 5~GHz residual map. The latter reveals "S"-shaped radio emission detected at high significance and which is rotationally symmetric about the flat-spectrum nuclei (a larger version is shown in Fig. 5). The 5~GHz inner-jet axis is offset by $\sim$45~degrees from the position angle of the vector between J1502SE and J1502SW (see Fig.~6). The location and close proximity of the two flat-spectrum nuclei support the view that the "S"-shaped radio emission is a modulation (or bending) of one pair of jets associated with one of the nuclear components. This type of "S-shaped" structure is commonly attributed to precessing jets, with a resulting radio-jet morphology similar to the famous X-ray binary SS\,433\cite{Hjellming1981}. Jet precession in AGN has long been predicted to be associated with the presence of binary black holes\cite{Begelman1980}. The spatially-resolved detection of the inner binary centred on the "S"-shaped jet emission provides, for the first time, direct evidence of this prediction. It therefore demonstrates that this is a promising method to find close-pair binary SMBHs that cannot be spatially resolved by current instruments ($<< 1$~milli-arcsecond angular resolution), yielding excellent targets for pointed gravitational wave experiments. The configuration where a primary black hole emits significant extended jet emission while the secondary appears to have none is also seen in the lowest separation binary SMBH known, namely the VLBI-discovered system 0402+379 at $z \sim 0.06$\cite{Rodriguez2006}. 

The host galaxies associated with J1502P and J1502S have elliptical galaxy morphologies and stellar masses of $M_{\rm *} = 1.7~\pm~0.3 \times 10^{11}$~M$_{\odot}$ and $2.4 \pm 0.4 \times 10^{11}$~M$_{\odot}$ respectively\cite{Fu2011b}. J1502P has a measured black hole mass of $\log{(M_{\rm BH}/\mathrm{M}_\odot)}$~$= 8.06 \pm 0.24$\cite{Fu2011b}, which is consistent with the black hole to bulge mass ($M_{\rm BH}-M_{\rm bulge}$) relation\cite{Haring2004}. If the same is true for J1502S, this implies that J1502SE/W both have black hole masses of $\sim$10$^8$~M$_\odot$ and each have a sphere of influence of $\sim$10~pc (the radius at which the black hole dominates the gravitational potential relative to the host galaxy). The J1502S optical spectrum reveals a single [OIII] component, despite the double flat-spectrum cores. If a circular orbit of radius $a= 138/2 = 69$~parsec and a black hole binary mass of $M_{12} = 2 \times 10^8$~M$_\odot$ are assumed, the maximum expected velocity offset between J1502SE/W is $V_{\rm J1502E/W} = \sqrt{{\rm G}M_{12} / a} \approx 110$~km\,s$^{-1}$ which is within the measured J1502S [OIII] line width of 384~km\,s$^{-1}$. Given that the original J1502+1115 selection was based on double-peaked [OIII] lines separated by 657 km\,s$^{-1}$, it is clear that neither the double-peaked [OIII] or the double-cored near-infrared imaging played any part in the selection of the much closer binary SMBH within J1502S. This implies that the only major selection criteria {\it for the close-pair binary within J1502S} are {\sl K}-band magnitude ({\sl K} $\sim$ 14-16); redshift ($z \sim 0.1-0.4$); and integrated 1.4~GHz flux density ($S_{1.4} \sim 1-60$~mJy). Based on simulations for radio continuum sources\cite{Wilman2008}, we expect a sky density of 5 such galaxies per square degree. Given that one sub-kpc binary was found from targeting six objects, this implies a sky density of $\Phi = 0.8 \pm ^{1.9}_{0.7}$~deg$^{-2}$ for the given selection criteria, where we quote the formal Poisson 1$\sigma$ uncertainties. Future VLBI surveys will be required to confirm this, given that the estimate is made from just six objects. While previous VLBI observations find a small fraction of dual/binary AGN\cite{Burke-Spolaor 2011}, they typically target bright sources which are usually associated with the most massive black holes ($M_{\rm BH} \gtrsim 10^9~\mathrm{M}_\odot$) for which there is a low probability of a similarly massive companion black hole. 

To date, there are very few confirmed dual/binary AGN, despite the expectations from our understanding of hierarchical structure formation through galaxy merging, combined with the observational evidence that every massive galaxy hosts a central SMBH\cite{Kormendy1995}. Of the dual/binary AGN with direct imaging confirmation, only three have sub-kpc projected separations and these systems are all below $z \lesssim 0.06$\cite{Komossa2003,Rodriguez2006,Fabbiano2011}. In addition to these, there are a handful of candidate close-pair binaries based on light curve analyses, double-peaked broad lines and radio jet morphology\cite{Valtonen2008,Boroson2009,Kaastra1992}, but these are not spatially resolved by current instruments. In Fig.~2 we show the projected spatial separations of the dual/binary AGN against redshift for a sample of sources that have been confirmed or discovered by spatially-resolved X-ray, optical/infrared or radio imaging. While there may be a degree of subjectivity on which sources should appear in this list, it illustrates that the two VLBI-discovered systems have over an order of magnitude lower spatial separations relative to X-ray and optical/infrared observations at similar redshifts. The J1502SE/W binary has a period of $P_{\rm bin} \sim 4$~million years, assuming a circular orbit and major axis of 138~parsec. This period is comparable to the lifetime of a typical radio source ($\sim$10 million years).

Limited observational constraints exist on the abundance of triple black hole systems. There are only five triple AGN candidates with their lowest ($a_1$) and second lowest ($a_2$) projected separations below 10~kpc (\emph{sub-10kpc systems} hereafter), which is the approximate effective radius for an elliptical galaxy. Fig.~3 shows the lowest and second lowest projected separations for all the candidate triple AGN systems found in the literature (8 in total for separations $\lesssim 100$~kpc). Six of the eight known triple systems (and all the sub-10kpc systems) have associated radio emission with 1.4 GHz luminosities that range from $\log{(L_{1.4}/\mathrm{W\,Hz^{-1}})} \sim 22 -25 $. Like J1502+1115, some of these triple systems have radio emission associated with multiple nuclei. From the literature and archival data we find that at least 9/15 ($>$60~percent) of the sub-10kpc SMBHs  have associated radio emission. This is a lower limit for two reasons: (1) sensitivity of the available observations; and (2) some of the radio imaging available does not have sufficient angular resolution to resolve all three galaxies or nuclear cores in a triple system. The fraction of triple AGN with associated radio emission is significantly higher than the typical value of $\sim$10 percent for AGN\cite{Jiang2007}, suggesting that triple systems lead to higher accretion activity and consequently a higher chance of jet triggering. None of the sub-10kpc systems were selected based on their radio emission, implying no clear biasing selection effects (see details of host galaxies in Methods section). Evidence for enhanced accretion has also been found for $\sim$10-100~kpc separation, hard-X-ray-selected dual AGN at low redshift\cite{Koss2012}. This is consistent with hydrodynamical simulations that find peak accretion and AGN luminosity occurs at the smallest separations ($<$1-10~kpc)\cite{VanWassenhove2012}. The above findings are therefore consistent with the expected disruption to gas dynamics caused by multiple black hole orbits. 

Multiple SMBH in-spiral and recoil rates, as well as the resulting orbital eccentricities are predicted to dominate the stochastic gravitational wave background spectrum in the nHz-$\mu$Hz range\cite{Hoffman2007,Blecha2011}. Therefore, constraints on the number density of triple systems is important to refine the strategy of the future pulsar timing array experiments. This discovery demonstrates that high angular resolution radio observations have a unique advantage in searching for multiple black hole systems over large cosmological volumes, and therefore predicts that radio facilities like the Square Kilometre Array (as a standalone array or part of a VLBI network) will lead to a dramatic increase in the number of known low separation ($<<$1~kpc) systems, as well as unresolved candidates for pointed gravitational wave experiments.

\section*{METHODS SUMMARY}

J1502+1115 was observed at 1.7 and 5 GHz with the European VLBI Network. The 305 metre Arecibo observatory was included in the 1.7~GHz observation. The 5 GHz observation used the e-VLBI technique, which performs real-time correlation. The correlation averaging time was 2~s and 4~s for the 5 GHz and 1.7 GHz visibilities respectively and we used 32 delay steps (lags) per sub-band. There were 4$\times$8 MHz sub-bands observed in both in both left and right circular polarizations with 2-bit sampling. Both observations used 4C39.25 as a fringe-finding source; and the nearby (angular distance $\Delta \theta \sim 51$~arcmin) source J1504+1029 as a phase calibrator. Both maps have a brightness scale uncertainty of $\sim$10~percent. The VLBI observations at 1.7 and 5~GHz were taken 18 months apart (and 23 months for the Arecibo run). The 5 GHz JVLA observations were performed in A-configuration and all details thereof are described in the literature\cite{Fu2011b}. We re-calibrated these data using the CASA software package and achieved flux densities consistent with previously published values\cite{Fu2011b}, within the uncertainties. The residual map shown in Fig.~1c was generated by taking the median of 500 realisations of maps made by subtracting point sources {\sl directly from the JVLA 5~GHz visibilities}. The point source positions and flux densities were derived from the VLBI 5~GHz components and the Briggs-weighted (robust=0) 5~GHz map for the J1502S and J1502P components respectively. Each realisation added a random offset in flux density and position, defined by their 1$\sigma$ uncertainties and drawn from a Gaussian distribution, so that the effect of calibration and deconvolution errors on the residual structure could be constrained. The 5~GHz VLBI observation was made 11 weeks prior to the JVLA 5~GHz observation\cite{Fu2011b}, limiting the uncertainty due to source variability in the comparison of the two.

\begin{addendum}
\item We thank John Magorrian, Aris Karastergiou, Scott Ransom and Bernie Fanaroff for very useful discussions. The European VLBI Network is a joint facility of European, Chinese, South African and other radio astronomy institutes funded by their national research councils. e-VLBI research infrastructure in Europe was supported by the European Union's Seventh Framework Programme (FP7/2007-2013) under grant agreement number RI-261525 NEXPReS. The financial assistance of the South African SKA Project (SKA SA) towards this research is hereby acknowledged. Opinions expressed and conclusions arrived at are those of the author and are not necessarily to be attributed to the SKA SA. www.ska.ac.za. Z.P. and S.F. acknowledge funding from the Hungarian Scientific Research Fund (OTKA 104539). Z.P. is grateful for funding support from the International Space Science Institute. 

\item[Contributions] R.P.D. was PI of the project and wrote the paper. Z.P. helped design, schedule, observe and calibrate the EVN observations as well as interpret the results. M.J.J. helped interpret the cosmological and astrophysical importance of the discovery and contributed significantly to the text. M.C. and R.P.F. helped in the binary SMBH interpretation and the physics thereof. S.F. calibrated the 1.7~GHz EVN observation and helped in the VLBI component interpretation. G.B. and I.H. helped with the technical aspects of the JVLA re-analysis, and the subtle interferometric effects at play. H.R.K. helped in the VLBI and GMRT proposals and wrote the software to calibrate the GMRT observation. K.G. and C.R. observed and calibrated the 16~GHz AMI observations. All authors contributed to the analysis and text. 

\item[Competing Interests] The authors declare that they have no
competing financial interests.
\item[Correspondence] Correspondence and requests for materials should
be addressed to Roger Deane (email: roger.deane@ast.uct.ac.za).
\end{addendum}

\newpage

\begin{figure}
	\includegraphics[width=0.95\textwidth]{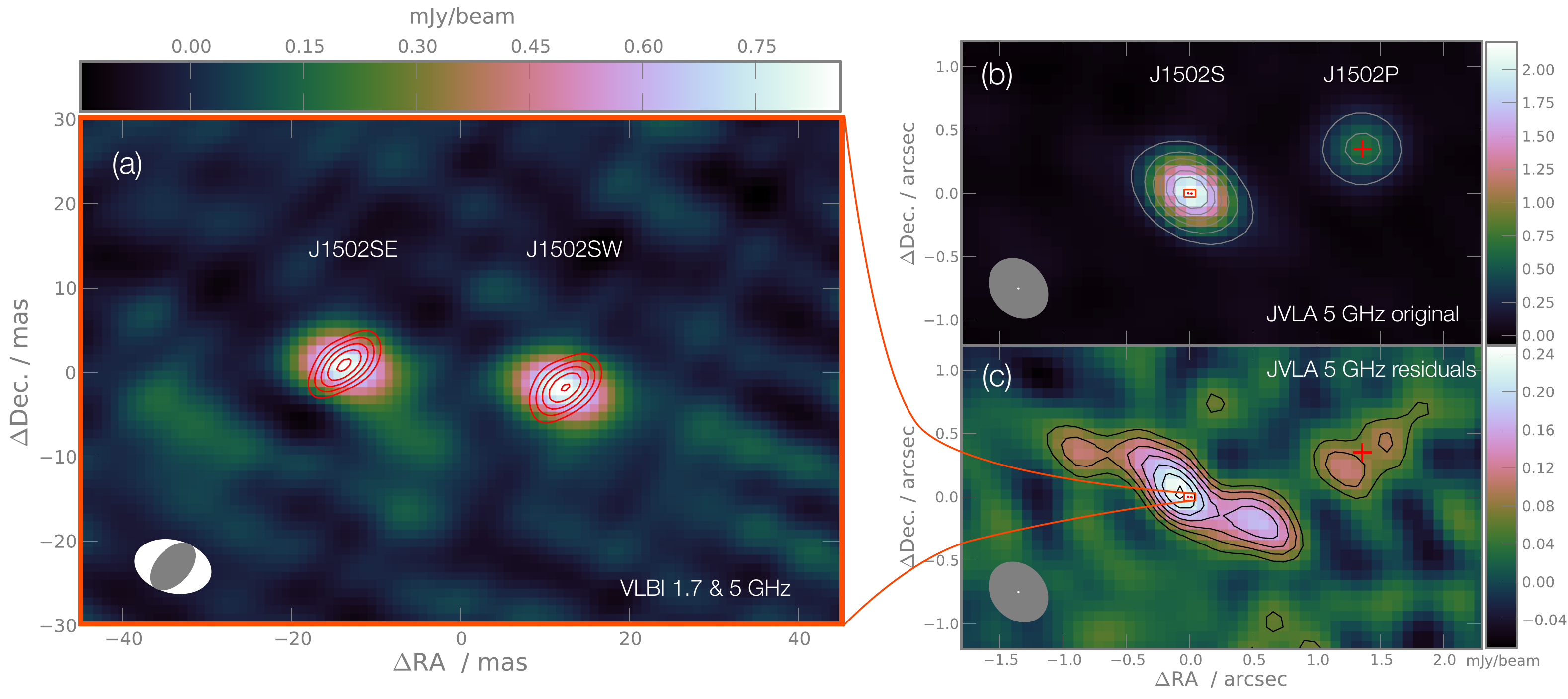}
	\caption{ {\bf VLBI and JVLA maps of the triple supermassive black hole system in J1502+1115.} {\bf (a) }5~GHz VLBI map (colour scale) showing the two components J1502SE and J1502SW. Over-plotted are red contours of the 1.7~GHz VLBI map with levels beginning at 400 $\mu$Jy\,beam$^{-1}$ and increasing in steps of 100 $\mu$Jy\,beam$^{-1}$ ($\sim$3$\sigma$). {\bf (b)} JVLA 5~GHz map imaged with Briggs {\sl uv}-weighting (robust=1). Over-plotted are the 5~GHz contours (grey) starting at 250~$\mu$Jy\,beam$^{-1}$ and increasing in steps of 300~$\mu$Jy\,beam$^{-1}$ (20$\sigma$). {\bf (c)} Point-source subtracted JVLA 5~GHz residuals revealing a rotationally symmetric "S"-shaped radio emission centred on VLBI components (within the red rectangle). There also appear to be small-scale jets centred on the J1502P radio nucleus (red cross). Over-plotted are the JVLA 5~GHz contours (black) starting at 60~$\mu$Jy\,beam$^{-1}$ (4$\sigma$) and increasing in steps of 2$\sigma$. Also shown are the frame boundaries of the left panel. Negative contours are also shown at -4$\sigma$. The white ellipses in lower left of all three panels indicate the point spread function FWHM of the 5~GHz VLBI map. The grey ellipse in the left panel shows the 1.7~GHz VLBI PSF, while the larger grey ellipses in the right hand panels show the JVLA 5~GHz PSF. } 
\end{figure}

\newpage 

\begin{figure} 
\centering
\includegraphics[width=0.95\textwidth]{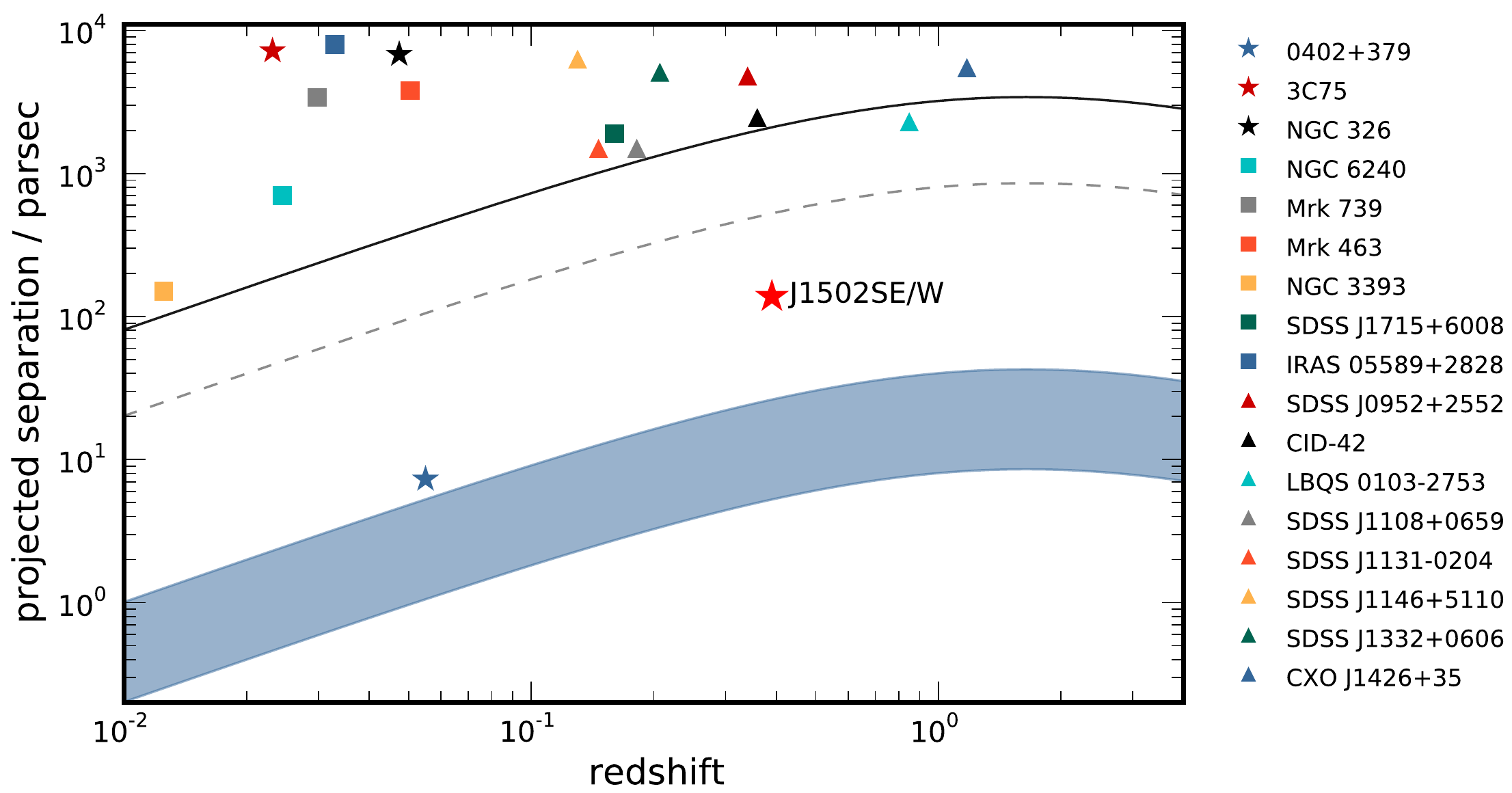}
\caption{ {\bf Dual/binary AGN confirmed or discovered with direct imaging.} Sample of dual/binary AGN confirmed or discovered by direct imaging at X-ray (squares), optical/infrared (triangles) and radio (stars) wavelengths. Typical spatial resolution limits of Chandra, the Hubble Space Telescope (HST) and VLBI are indicated by the solid and dashed lines, and shaded blue area respectively. This illustrates that high angular resolution radio observations are able to survey a significantly larger cosmological volume for binary AGN with separations comparable to the black hole gravitational spheres of influence ($<$10~pc for $M_{\rm BH} \sim 10^8$~M$_{\odot}$).}
\end{figure}

\begin{figure}
\centering
\includegraphics[width=0.95\textwidth]{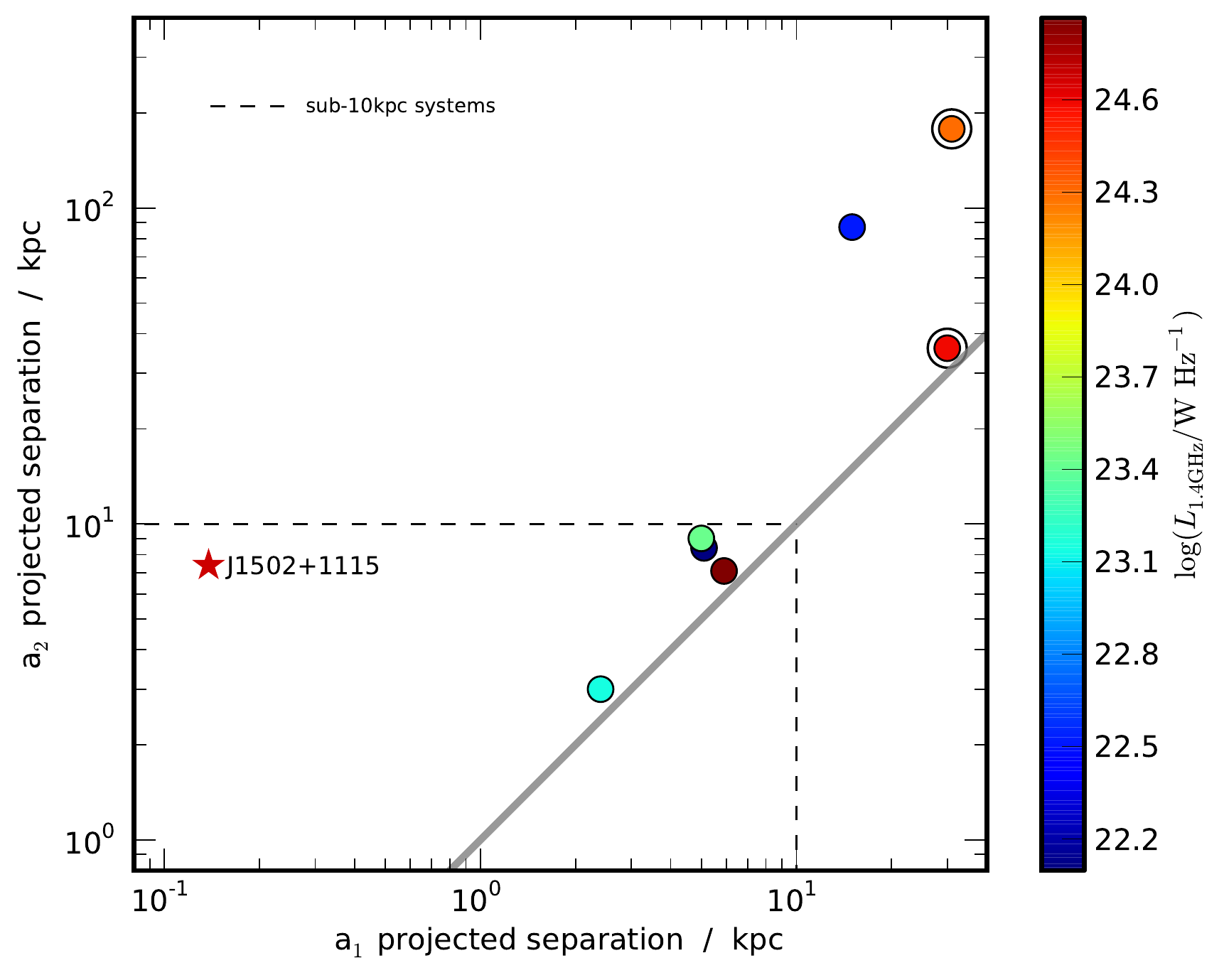} 
\caption{ {\bf Projected separations of candidate triple active galactic nuclei.} The axes denote the lowest and 2$^{\rm nd}$ lowest separations between AGN in the list of candidate triple AGN systems. The solid black line indicates a ratio of unity. The dashed lines isolate the sub-10kpc systems (see text). The colors correspond to the total integrated radio luminosity of the AGN plus the host galaxy, since low angular resolution limits a fair AGN radio luminosity comparison. Nonetheless, the plot supports enhanced accretion within triple systems, particularly in the sub-10kpc systems. J1502+1115 is an isolated region in the parameter space, likely due to the bias against low separation for low resolution optical/near-infrared and X-ray observations. The two black circled points (which are triple quasar systems) indicate upper radio luminosity limits from archival observations. The triple AGN candidates and their references are listed in the Methods section. Note that all sub-10kpc systems are equidistant within a factor of 2, except for J1502+1115. }
\end{figure}

\vspace{30 mm}

\newpage

\begin{methods}

\section{Radio observations}

\subsection{e-VLBI 5 GHz. } 

J1502+1115 was observed at 5 GHz with the European VLBI Network using the e-VLBI technique on 12 April 2011 for approximately 1.5 hours. The participating telescopes were Effelsberg (Germany), Jodrell Bank (Mk II, UK), Hartebeesthoek (South Africa), Sheshan (China), Medicina (Italy), Onsala (Sweden), Toru\'n (Poland), Yebes (Spain) and the phased-array Westerbork Synthesis Radio Telescope (WSRT, Netherlands). The data were streamed real-time to the EVN Data Processor at the Joint Institute for VLBI in Europe (JIVE) at a 1024 Mbps data rate. The correlation averaging time was 2 s and we used 32 delay steps (lags) per sub-band. The coordinates used for correlation was the J1502+1115 1.4~GHz centroid determined from the Faint Images of the Radio Sky at Twenty-Centimeters (FIRST) survey\cite{Becker1995}. J1502SE/W are detected sufficiently near the phase centre ($\Delta \theta \lesssim 1.3$~arcseconds) that time and bandwidth smearing losses are negligible. There were 4$\times$8 MHz sub-bands observed in both LCP and RCP polarizations with 2-bit sampling. The initial clock searching was carried out before the experiment using the fringe-finder source 4C39.25. J1502+1115 was phase-referenced to the nearby ($\Delta \theta \sim 51$~arcmin) calibrator J1504+1029 for 1.3 minutes per 5 minute calibrator-target cycle. The J2000 coordinates of the reference source (RA=15h 04m 24.979782s, Dec=+10d 29' 39.19858") were taken from the Very Long Baseline Array (VLBA) Calibrator Survey (http://www.vlba.nrao.edu/astro/calib/index.shtml) and have a positional uncertainty of $\sim$0.02 milli-arcseconds. The resultant map has a Briggs-weighted (robust=0) synthesized beam with a 9.1$\times$6.1~milli-arcseconds$^2$ FWHM at a position angle of 76 degrees east of north. The r.m.s. noise in the Briggs-weighted (robust=0) 5~GHz map is $\sim36 \, \mu$Jy\,beam$^{-1}$ and the flux scale uncertainty is $\sim$10~percent. The J1502P component is not detected in the VLBI observations with a 5$\sigma$ upper limit of $S_{\rm 5GHz} \lesssim 180~\mu$Jy\,beam$^{-1}$. Components J1502SE and J1502SW are detected with flux densities of $S_{\rm 5 GHz} = 857 \pm 49$ and $872 \pm 49 \ \mu$Jy respectively, where the uncertainties are the quadrature sum of the fitting uncertainty and the r.m.s of the map measured $\sim$2 arc seconds from the phase centre. This e-VLBI observation was made 11 weeks prior to the JVLA 5~GHz observation\cite{Fu2011b}, limiting the uncertainty due to source variability in the comparison of the two.

\subsection{VLBI 1.7 GHz. } 
Based on the detection of two compact components in the e-VLBI 5~GHz map, we observed J1502S at 1.7~GHz using the European VLBI Network. The observations were performed on the 24 October 2012 and 13 March 2013. Antennas included in the VLBI array for the 2012 run were the same as the 5~GHz e-VLBI observations, as well as Urumqi (China) and the Russian KVAZAR Network stations Svetloe,  Zelenchukskaya and Badary. The second epoch observations included the following antennas: Arecibo, Effelsberg, Jodrell Bank, Onsala, Medicina, Toru\'n, Noto and WSRT. J1502S was observed for  approximately 3 hours in total. The data were 2-bit sampled. The data in this case were recorded with Mark5As and correlated at a later time at JIVE. There were 128 delay steps per sub-band and the integration time was 4 seconds. The same fringe-finder, phase reference target and duty cycle was used as in the e-VLBI observation.  The coordinates used for correlation were determined from the e-VLBI 5~GHz detection. The 1.7 and 5~GHz VLBI centroids are consistent within the fitting uncertainties ($\lesssim$1.5~mas, 7.5~parsec) for non-self-calibrated 1.7 GHz \emph{uv}-data. Including the Arecibo baselines in the 1.7~GHz \emph{uv}-data and self-calibrating all Arecibo baselines leads to a higher angular resolution and fidelity map, but with compromised absolute astrometry. The resultant offset is ($\Delta$RA, $\Delta$Dec.) = (0.55~mas, 3.15~mas). This shift is corrected for in the map shown of J1502SE/W in the left panel of Fig.~1 of the main paper. The measured flux densities for self-calibrated and un-self-calibrated data are consistent within the fitting uncertainties. The naturally-weighted map has an r.m.s  of $\sim31 \, \mu$Jy\,beam$^{-1}$ and a synthesized beam FWHM of 6.5$\times$3.9~milli-arcseconds$^2$  at a position angle of -42 degrees east of north. The J1502P component is not detected in the VLBI observations with a 5$\sigma$ upper limit of $S_{\rm 1.7GHz} \lesssim 170~\mu$Jy\,beam$^{-1}$. Components J1502SE and J1502SW are detected with flux densities of $S_{\rm 1.7 GHz} = 954 \pm 50$ and $920 \pm 49 \ \mu$Jy respectively, where the uncertainties are the quadrature sum of the fitting uncertainty and the r.m.s of the map measured $\sim$2 arc seconds from the phase centre. The VLBI observations at 1.7 and 5~GHz were taken 18 months apart (and 23 months for the Arecibo run), yielding an  apparent proper motion upper limit of $\Delta \mu \lesssim 5$~parsec\,yr$^{-1}$.

\subsection{Arcminute Microkelvin Imager (AMI) 16 GHz. }

J1502+1115 was observed on 6 December 2012 for 12 hours by the AMI Large Array\cite{Zwart2008} for approximately 1.5 hours. The telescope observes between 13.5 and 18~GHz in six spectral
channels of 0.75~GHz bandwidth. The source J1504+1029 was observed for 100~s every 600~s for accurate phase calibration. 
Flux density calibration was performed using short observations
of 3C\,286 which was calibrated against VLA measurements\cite{Perley2013}. The data were flagged and calibrated using {\sc reduce}, a local software
package developed for AMI. The calibrated data were then mapped and {\sc clean}ed in {\sc aips} using the {\sc imagr} task.  The integrated flux density of the detection was
$S_{\rm int} \, \sim \,$1.17~mJy. The final thermal noise on the continuum map was $35 \, \mu$Jy\,beam$^{-1}$ and a flux scale uncertainty of 5~percent. The map has not been corrected for the telescope primary beam response, which is well modelled by a Gaussian of FWHM 5.6~arcminutes. The {\sc clean} restoring beam has a FWHM of $64 \, \times \,  23$~arcseconds at a position angle of 9~degrees. 
We do not expect confusion with other sources within this beam based on the higher resolution and sensitivity JVLA continuum 8~GHz map.

\subsection{Giant Metrewave Radio Telescope (GMRT) 610 MHz. }
J1502+1115 was observed with the GMRT array on 17 December 2012 at 610~MHz for approximately 45 minutes. J1502+1115 was phase referenced to the nearby ($\Delta \theta \sim 51$~arcmin) calibrator J1504+1029. Amplitude calibration was performed using 3C\,286. Data processing was performed with an automated calibration and imaging pipeline based on {\sc python, aips} and {\sc parseltongue}\cite{Kettenis2006} which has been specially developed to process GMRT data. The full technical description of this calibration and imaging pipeline is extensive and is presented in detail elsewhere\cite{Mauch2013}. The integrated flux density of J1502+1115 at 610~MHz is $S_{\rm int}$ = 22.84$\pm$0.95~mJy, where the uncertainty is the quadrature sum of the fitting uncertainty and the local noise near the J1502+1115 detection. The {\sc clean} restoring beam is $5.5 \, \times \,  4.4$~arcseconds at a position angle of 59~degrees, and therefore does not spatially resolve J1502P and J1502S.

\subsection{JVLA}

The JVLA observations were performed on 29 June 2011 at 1.4 and 5~GHz. The 8 GHz observations were performed on 14 June 2011. All three of these bands were observed while the JVLA was in A-configuration. The details of the observations are reported in the literature\cite{Fu2011b}. We re-calibrated all three observations using the CASA software package and achieved flux densities consistent with previously determined values\cite{Fu2011b}, within the uncertainties.

\section{Radio spectra of radio components}

In Fig.~4 we show the radio spectra of the various components. The component J1502P has a steep spectrum and is unlikely to contribute to the measured 15.7 GHz AMI flux density. It therefore appears that the AMI map is dominated by the flat-spectrum cores.

\section{JVLA 5~GHz residual map}

The residual map shown in Fig.~1c was generated by taking the median of 500 realisations of maps made by subtracting point sources {\sl directly from the JVLA 5~GHz visibilities}. The point source positions and flux densities were derived from the VLBI 5~GHz components (observed 11 weeks prior to the JVLA 5~GHz observation) and the Briggs-weighted (robust=0) 5~GHz map of the JVLA 5~GHz data itself for the J1502S and J1502P components respectively. Each realisation added a random offset in flux density and position, defined by their 1$\sigma$ uncertainties and drawn from a Gaussian distribution, so that the effect of calibration and deconvolution errors on the residual structure could be constrained. Note the JVLA 8~GHz residual map shows similar structure, but at a lower signal-to-noise ratio.This analysis was performed using the MeqTrees software package\cite{Noordam2010} (www.meqtrees.net). The JVLA-detected jet emission is not detected in the VLBI maps, due to insufficient brightness temperature sensitivity.

\section{JVLA 5~GHz inner jet}

The JVLA residual map discussed above (and shown Fig.~1c) was generated using Briggs weighing (robust=1). In Fig.~6, we show the same JVLA 5~GHz residuals, however with Briggs weighting (robust=0). This results in higher angular resolution (synthesized beam FWHM = 0.39$\times$0.32~arcsec$^2$, position angle = 45 degrees east of north), but with poorer sensitivity (1$\sigma \sim 30\ \mu$Jy\,beam$^{-1}$), or more importantly, poorer brightness temperature sensitivity. This results in a map that traces the higher brightness temperature emission which we use to probe the central region of J1502S (see Fig.~6). This figure shows the JVLA~5~GHz inner jet at a position angle of $\sim$45 degrees east of north. This is significantly mis-aligned with the position angle of the vector between J1502SE and J1502SW ($\theta_{\rm J1502SE/W} \sim 87$ degrees). J1502SE is within $\sim$20~milli-arcseconds of the brightest pixel in the JVLA 5~GHz residual map, well within the absolute JVLA astrometry of 100~milli-arcseconds\cite{Ulvestad2009}. This supports the view that the "S"-shaped jet emission seen in Fig.~1c stems from one of the two VLBI-detected cores. The large mis-alignment is inconsistent with any scenario where the two VLBI components are interpreted as young hotspots stemming from a single, undetected core. Furthermore, the straight north-east inner jet with projected length $\sim$0.4~arcseconds ($\sim$2.1~kpc) and a jet axis position angle of $\sim$45~degrees is also inconsistent with a scenario where one of the VLBI components is an AGN core and the other a hotspot along the jet trajectory. A detailed discussion is also included in the following section.


\section{Alternative interpretations of the parsec-scale radio emission in J1502S}

In the two sub-sections below we distinguish between jets and hotspots to simplify comparison to the literature.

{\it \noindent Double-jet or core-jet scenarios}\newline
A possible interpretation of the double flat-spectrum structure of J1502S
is that we observe the synchrotron self-absorbed bases of a pair of oppositely
directed jets from a single AGN. Similar structures have been seen in
nearby type-II AGN, as well as in the Galactic X-ray binary system SS\,433\cite{Paragi1999}.
We may exclude this scenario for a number of reasons. 

\begin{enumerate}
\item We do not see elongated core-jet structure in either of the two components, which is contrary to what is typical of self-absorbed AGN cores. 
\item We would not expect these to be unresolved at VLBI scales given the fact that the two components are separated by 26~milli-arcseconds, and the separation between the two jet sides should be comparable to the size of the core itself\cite{Blandford1979}. 
\item The 138~pc physical separation between the two components is an order of magnitude larger than the typical size of radio cores ($\sim$1-10 parsec) seen in the brightest AGN\cite{Lobanov1998,Hada2011}.
\end{enumerate}

{\it \noindent Double-hotspot or core-hotspot scenarios} \newline
There are a number of arguments against the scenario where the two compact J1502SE/W components are hotspots as observed in the class of compact symmetric objects (CSOs); or alternatively that one component is a hotspot and the other the core. We outline the most important points below. 
 
\begin{enumerate}
\item Both components have flat spectra, which is uncharacteristic of hotspots which have typical spectral indices of $\alpha_{\rm hotspots} < -0.5$\cite{An2012}. Spectral indices are even steeper for CSOs with a projected linear size (hotspot separation) greater than $\sim$100~pc\cite{An2012}. 
\item The population of CSOs have typical luminosities of 10$^{27}$~W\,Hz$^{-1}$, which is 4 orders of magnitude larger than J1502SE/W\cite{Odea1998,An2012}. While this may be a selection effect, the lowest luminosity CSO that we have found in the literature has a 1.4 GHz radio luminosity that is 1 order of magnitude larger than that of J1502SE/W\cite{An2012}. 

\item Similar to the double-jet/core-jet scenario discussed in the previous section, higher frequency observations probe nearer the base of the jet since it becomes optically thin closer to the central black hole\cite{Lobanov1998,Hada2011}. In the core-hotspot scenario where the distance is as large as 138~pc, this would result in a observable frequency dependent core-shift (i.e. decreasing separation between the two components) which we do not see: the separation of J1502SE and J1502SW appears independent of observing frequency.

\item In the double-hotspot scenario, there is no core emission detected (to $L_{\rm 1.7GHz}  \sim 10^{22}$~W\,Hz$^{-1}$). 

\item The straight $\sim$2 kpc long inner jet of J1502S has a 45 degree misalignment with the vector between J1502SE/W. This would imply a dramatic, almost instantaneous change in angular momentum.

\item Assuming that one or both J1502SE/W are young hotspots; this would imply that we happened to observe J1502+1115 within a very short time period since this dramatic change in angular momentum. If we assume that the jet axis is in the plane of the sky and a very conservative jet speed of 0.01-0.1c, the shift in momentum must have occurred within the last $2 - 20 \times 10^3$~years. For a nominal radio source lifetime of $\sim$10$^7$~years, this represents a 0.02 - 0.2 percent probability. Indeed, hotspot velocities from young radio sources are typically 0.3-0.5c, and almost never below 0.1c, corroborating our claim that this is a very conservative estimate\cite{An2012}. 
\end{enumerate}

So in summary, in order for a core-hotspot or double-hotspot alternative to be true, the hotspot(s) would not only need to have highly uncharacteristic flat spectral indices (and potentially an order of magnitude larger core size compared to what is observed in the highest luminosity radio sources); but also to have been observed almost instantaneously after a major shift in angular momentum has occurred.  Although the binary SMBH interpretation may be considered a (hitherto) exotic one, combining the above arguments with rotationally-symmetric "S"-shaped flux strongly favours the binary SMBH interpretation.

\section{Triple AGN candidates}

The triple AGN candidates plotted in Fig.~3 of the main text and the relevant references are SDSS J1027+1749\cite{Liu2011}, NGC 3341\cite{Barth2008,Bianchi2013}, QQ 1429-008\cite{Djorgovski2007}, GOODS J123652.77+621354.7\cite{Schawinski2011}, QQQ J1519+0627\cite{Farina2013}, NGC 6166\cite{Tonry1984,Ge1994} and the NGC 835 group\cite{Koss2012}. The 1$\sigma$ upper limits on 1.4 GHz radio luminosity for the two triple quasars are determined from the local noise estimate around their respective sky coordinates in VLA FIRST survey\cite{Becker1995}. There is a broad range in redshifts for this list of triple AGN candidates spanning from $z \sim$~0.005 to $\sim$2. Primary host galaxy morphology does not appear to play a strong role in preferentially selecting the six triple AGN systems with radio emission (3~disks, 2 ellipticals, 1 cD).

\begin{figure}
\centering
\includegraphics[width=0.95\textwidth]{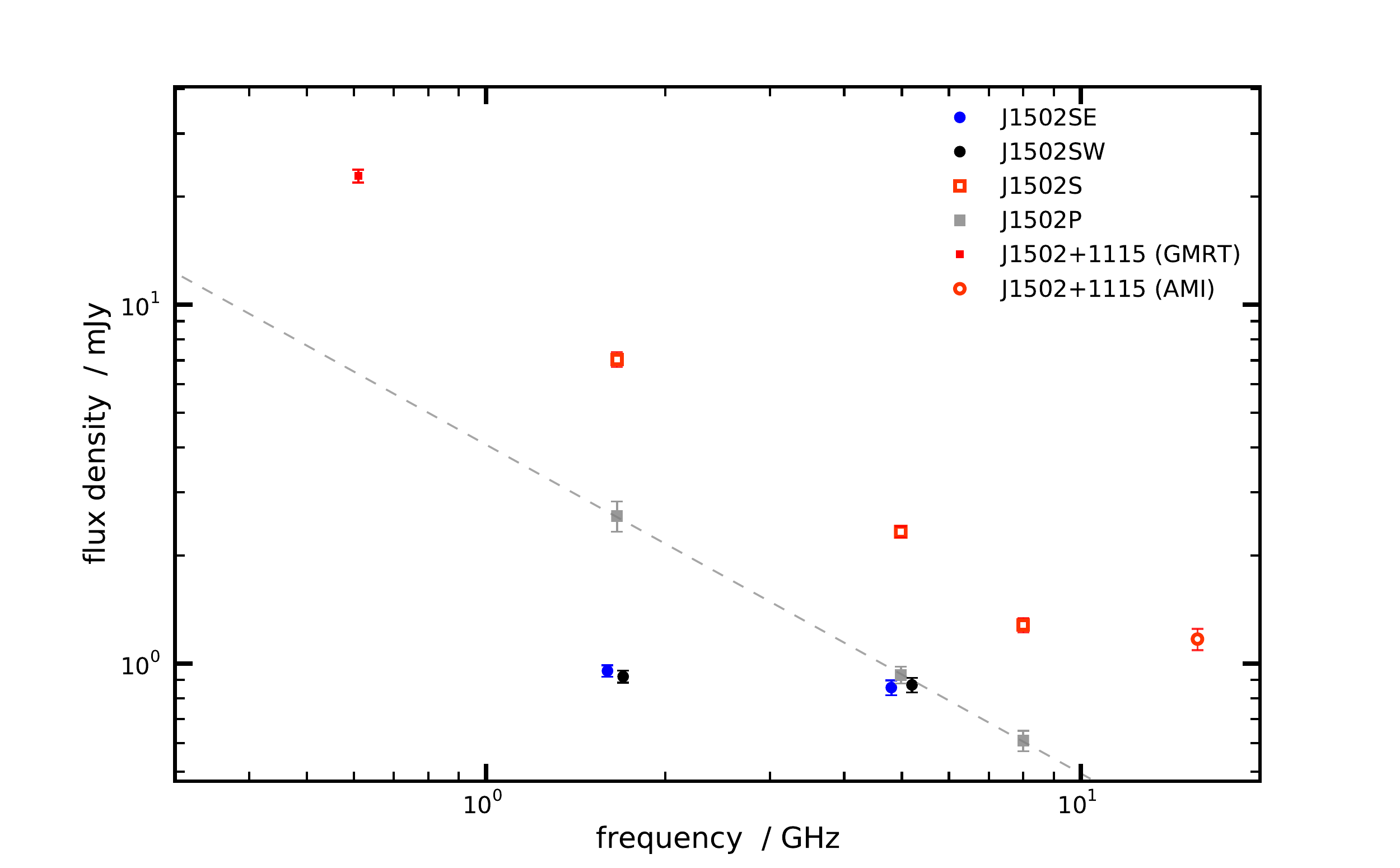} 
\caption{  {\bf Radio spectrum of the radio components in J1502+1115.}  Both J1502S (open red squares) and J1502P (grey squares) are steep-spectrum radio sources between 1.4 and 8 GHz, as measured by JVLA observations. The two flat spectrum cores (J1502SE/W, filled circles) are the likely cause of the flattened spectrum of J1502S at higher frequency, as measured by the AMI 15.7~GHz detection, labelled as J1502+1115 (AMI). The error bars represent the $\pm$1$\sigma$ standard deviation on the flux measurement. J1502+1115 (GMRT) indicates the 610 MHz detection with the GMRT. Note that J1502SE and J1502SW are marginally offset in frequency purely for clearer illustration.}
\end{figure}

\vspace{40 mm}

\begin{figure}
\centering
\includegraphics[width=0.95\textwidth]{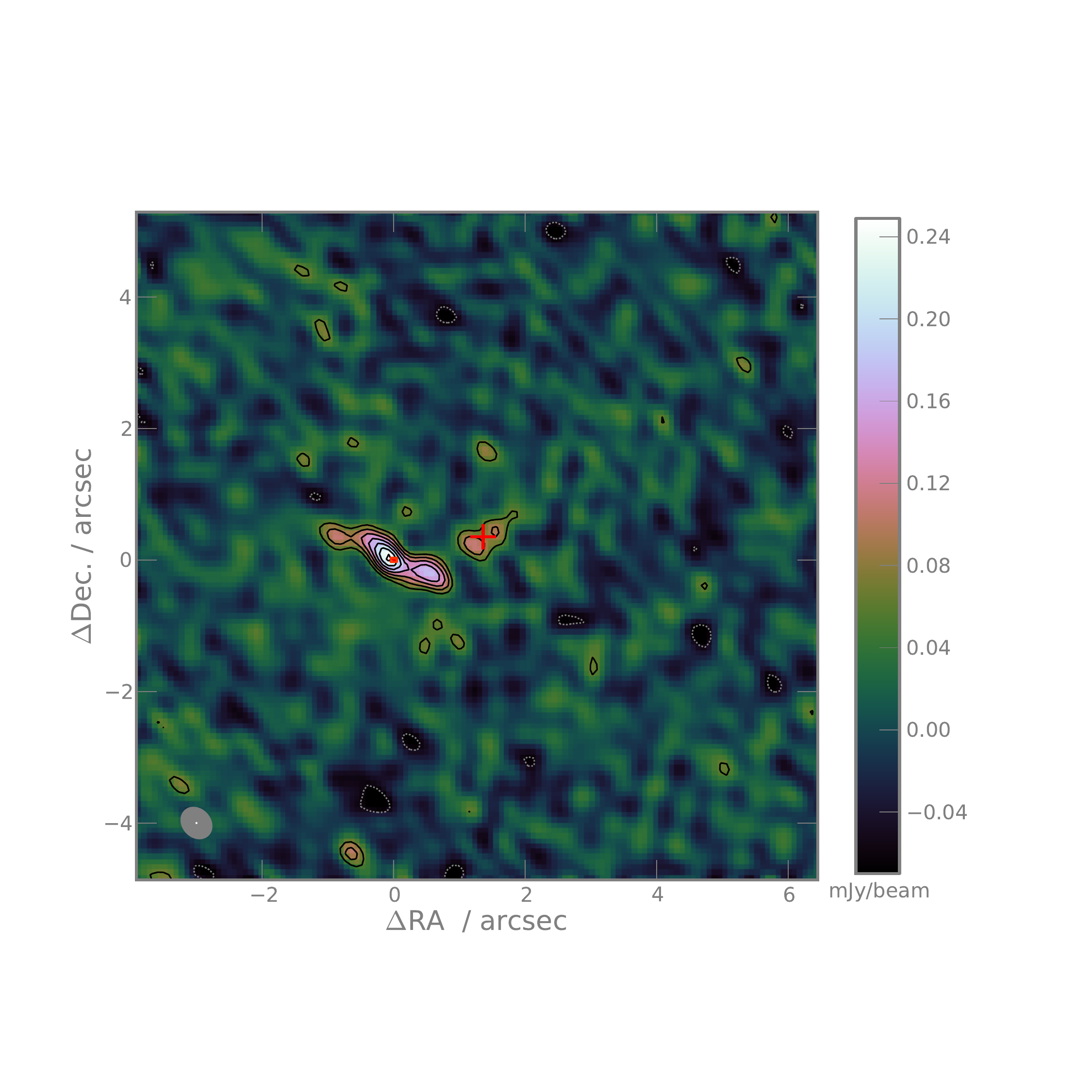} 
\caption{  {\bf Larger field-of-view JVLA 5 GHz map of J1502+1115 to demonstrate map fidelity.} The colour scale shows the same JVLA 5~GHz residuals shown in Fig.~1c but with the full 128$\times$128 pixel median map generated from the Monte Carlo realisations. The red square indicates the map boundary of the VLBI map shown in Fig.~1a. The red cross denotes the centroid of the point source subtracted J1502P component. Contour levels are the same as in Fig.~1c. The grey ellipse (lower left) represents the FWHM of the Briggs-weighted (robust=1) PSF, while the white dot shows the VLBI 5~GHz PSF. }
\end{figure}

\vspace{30 mm}

\begin{figure}
\centering
\includegraphics[width=0.95\textwidth]{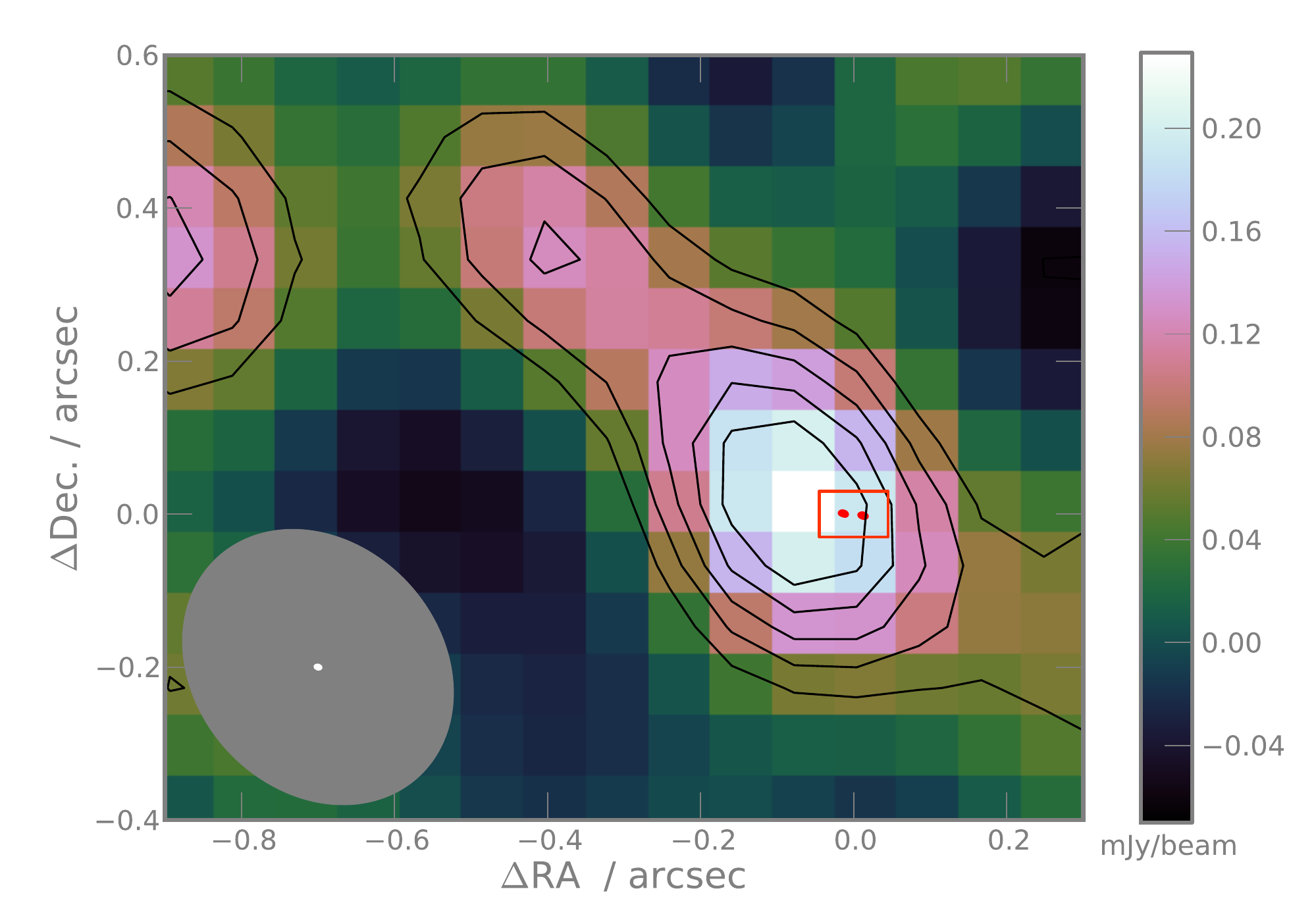} 
\caption{  {\bf Zoom-in of the high brightness temperature inner jet emission of J1502S.} The colour scale shows the same JVLA 5~GHz residuals shown in Fig.~1c but imaged with Briggs {\sl uv}-weighting (robust=0) to highlight the position angle of inner north-east J1502S jet. This is mis-aligned with the vector between J1502SE and J1502SW (red dots) by $\sim$45 degrees. The black JVLA 5 GHz contours start at 60 $\mu$Jy\,beam$^{-1}$ (~2$\sigma$) and increase in steps of 1$\sigma$. The grey ellipse (lower left) represents the FWHM of the Briggs-weighted (robust=0) PSF, while the white ellipse shows the VLBI 5~GHz PSF.  The red square indicates the map boundary of the VLBI map shown in Fig.~1a.}
\end{figure}

\end{methods}
\end{document}